# AN ECONOPHYSICS MODEL FOR THE MIGRATION PHENOMENA

Anca GHEORGHIU and Ion SPÂNULESCU[*]

***Abstract.*** *Knowing and modelling the migration phenomena and especially the social and economic consequences have a theoretical and practical importance, being related to their consequences for development, economic progress (or as appropriate, regression), environmental influences etc. One of the causes of migration, especially of the interregional and why not intercontinental, is that resources are unevenly distributed, and from the human perspective there are differences in culture, education, mentality, collective aspirations etc. This study proposes a new econophysics model for the migration phenomena.*

## 1. Introduction

Physics, is the most suitable for modelling the economic phenomena and structures or financial-banking operations, because it takes into consideration the process variables characteristics and permits to use some procedures – including the mathematical one – especially probability theory for minimizing or eliminating such influences depending on human factor and unexpected phenomena also, which cannot be predicted by direct methods.

Most econophysics approaches, models and papers that have been written so far refer to the economic problems including systems with a large number of elements such as financial or banking markets, stock markets, incomes, production or product's sales, individual incomes and some others models where statistical physics methods, Boltzmann, Gibbs and some other statistical distribution types are mainly applied [1÷14].

In this paper is proposed and analysed a new econophysics model for the migration phenomena.

Interregional, interstate or intercontinental migration phenomena, known since ancient times and is of particular importance both historically, politically, but mostly by the economic consequences which these masses movements have on their home regions or countries but also on the countries where these masses are moving sometimes forever. Knowing and modelling of the migration phenomena and especially the socio-economic consequences are important both theoretically and practically, being connected – as mentioned above – especially for their consequences on the economic evolution and progress (sometimes the regress.

---

[*] Hyperion University of Bucharest, 169 Calea Călăraşilor, Street, Bucharest, Romania, anca.gheorghiu@gmail.com.

## 2. Migration causes.
## The Classical Models for the Migration Phenomena

The mass migration causes, especially of the labour are multiple and extremely diversified, and the inventory and especially the analysis of this phenomenon would require too much space to be addressed in this paper, which generally refers to the modelling of the economic phenomena/processes, using patterns from physics and other sciences.

Excluding small reverse type migrations, which occur to the local border traffic or the international seasonal tourism (summer, winter), below we'll mention some of the types of migration, mainly economic or politic, and caused by natural disasters or catastrophes, wars, revolutions, conflicts etc. (e.g. Afghanistan, Iraq, Libya, Syria, Somalia, devastating Asian earthquakes etc.)

Excepting the great migrations in history (see the "barbarian" invasions) or these produced by major natural disasters or related to socio-political or military conflicts (dictatorships, revolutions, wars etc.) there are three major categories of migrations:
1. The migration of people acceding to a higher standard of living or want a better education or job change etc.
2. The migration of the population unemployed or unskilled or poorly qualified, which takes place mainly in periods of economic recession.
3. Seasonal migration, including of those who return to their home regions (reverse migration).

Regarding the last category (seasonal migration), it is linked to the performance of agricultural labours (crop pickers, "strawberry pickers" etc.) or to the achievement of large industrial objectives or buildings (highways, dams, forestry, residential neighbourhoods or towns, gigantic compound etc.).

Typically, these types of migration are small in number and also in space or time extent.

As well, linked to the first two categories of migration, people generally may migrate to find work, or to accede to a higher standard of living, with higher wages etc. These migrations occur in the rich regions (or countries), developed with public or social services, goods or abundance of job opportunities in the industrial centres, commercial services and access to higher education (universities, research institutes etc.).

In some regions (countries), sometimes there may be large masses of unemployed or unskilled workers, either because of demographic explosions (India, Pakistan etc.), either for economic reasons, as economic and financial crises, bankruptcies or heavy activity restrictions, elimination of large mines or industrial objectives, depleting the soil or subsoil (oil, coal, gold, forestry or agricultural etc.), either to poor economic development (underdeveloped countries) and standard of living (services, goods etc.).

Here can also be embedded the small migrations of political causes, such as those from ex-socialist countries or from some Afro-Asian countries with totalitarian regimes.

Another cause of migration, especially of those inter-regional or inter states, is that labour is not homogeneous in terms of qualifications, social status, human type etc. In highly developed states (USA, Germany, Japan etc.) migrate qualified individuals, sometimes with very high qualifications (programmers, teachers, scientists) as well as workers with low qualifications or no qualifications, for maintenance work, considered by locals as being below the national level of payment or difficult to accept for their status, as sweepers, scavengers, unskilled jobs in restaurants or shops, day labourers, shippers, vendors etc., for immigrants recruited from Asian, African countries etc., sometimes working without legal status.

Small migration may also occur if the policies of big companies are to open branches of production (cars, electronics or computer equipment, etc.) in other countries, or following the opening of large shipyards, mining or similar, as mentioned above.

The migrations of modern and contemporary era may have some positive effects, leading in some cases, to some smoothing or removal of disparities in employment of labour, payroll, development of regions (those who have not yet exploited) etc.). At times, some effects of migration may be less desirable or even disastrous, adding to an increase of underdevelopment and impoverishment in abandoned areas, which are deprived of human capital for economic development (labour).

The most damaging aspect – especially for countries losing skilled workforce – attracted by higher salaries and superior working conditions (research laboratories, documentation funding etc.) – is the migration of scientists from less developed countries (India, Pakistan, ex-socialist countries including Romania etc.) and other more developed countries (in Europe or Asia) in the United States of America, a phenomenon known as "brain drain". We'll not mention here the ethical aspects of the problem, especially as the losses from leaving country specialists, for whose training (schooling) countries have spent large sums of money, but we will reveal that there is an exodus and it can also be included in small migrations category but has major economic implications.

For the economic analysis and prognosis, particularly related to the labour migration, have been developed several models including different variables about the factors that determine migration and influence the economic development of the countries (regions), to absorb labour or, conversely, lead to economic regression of the countries that have lost a part of the workforce, particularly the medium or highly qualified (experts, scientists).

Classical models to describe the migration phenomena, mostly of economic type, were developed based on applying highly simplifying assumptions and therefore the results of their application or their performance was quite poor. These models focus mainly on differences in personal income (derived from wages or from other sources, e.g. businesses) between host regions (countries) and those from which migrants originated. But it must be specified – as has been said before – there are other causes of migration, in addition to these strictly connected to wages or income. Therefore, taking into account the many causes of migration, "classical" models have been corrected or substantially improved, and respectively, adjusted for specific causes groups or to gain even a global or general quality.

Such an "improved" model is based on consideration of the benefits of human capital; hence its emphasis is on migrant's interests and qualities. This model has the advantage that into the theory of migration based on that model can be incorporated the costs and benefits of migration by introducing the net migration from one region $i$ to another region $j$:

$$V_{ij} = R_{ij} - C_{ij} \qquad (1)$$

where:

$V_{ij}$ – represents the net present value of the migration potential from region $i$ to region $j$;

$R_{ij}$ – total benefits expected to be obtained (both monetary and otherwise);

$C_{ij}$ – current total value of the costs that may occur (both monetary and otherwise).

The economic, linguistic and cultural differences and the demographic dynamics between two regions are reflected in Figures 2 to 6 (see Appendix). The Figures refer to the price level indices and the relative dynamic of population according to, the gold reserves, the number of computers/1000 inhabitants and the number of internet users/1000 inhabitants.

The monetary flows migrations, the consumer goods migrations, the populations migrations etc. to the economic attraction centres, also called **economic attractors**, can be assimilated with gravitational or electrostatic attraction models.

The greater the distance to the attractor is, with the respect to time, and also to the transit difficulties or environmental, economic or political barriers, the greater the costs (consumption) attracted elements are.

The other migration model is based on the gravitational attraction principle, hence on the principles of physics (mechanics) in this case based on the law of attraction discovered by Newton. This model, called the gravitational model of migration, proposed and developed by geographers, has the general form [16,17]:

$$M_{ij} = f(A_{ij}, B_{ij}, D_{ij}) \qquad (2)$$

where:

$M_{ij}$ – is the total migration from region (country) $i$ to region (country) $j$;

$A_{ij}$ – a group of parameters of the migration flow from the country of origin (e.g. the population in the region or country of origin);

$B_{ij}$ – a group of specific variables of the destination country (region) (e.g. the population in the region or country of destination);

$D_{ij}$ – costs (losses) due to the distance between the two regions (transportation costs including obtaining information on the destination region, documents cost etc.)

For the gravitational model was taken into consideration the introduction of economic variables. Thus, an empiric model for determining the size of migration $M_{ij}$ (Total mass of migrants from region $i$ to region $j$) can be as [15,16]:

$$M_{ij} = f[P_i, P_j, D_{ij}, (U_j - U_i), (W_j, W_i)] \qquad (3)$$

where:

$P_i$ – represents the population of the region (country) $i$;

$P_j$ – the population of the country $j$;

$D_{ij}$ – the "economic" distance, i.e. transport costs between $i$ and $j$, documentation costs, papers costs etc.;

$U_i$ – the unemployment rate in the country (region) $i$;

$U_j$ – the unemployment rate in the country $j$;

$W_i$ – the wage rate in the country (region) $i$;

$W_j$ – the wage rate in the country (region) $j$.

In this paper we restrict the categories of migration for economic reasons in which the oriented mass movements of people take place to find better living conditions or to find work, or to raise their training level etc.

## 3. The Coulomb model for the migration phenomena for economic reasons

In the following we will refer in particular at the large-scale migration, especially interregional or even intercontinental such as those known as "migrations" from the depths of Asia to Europe (that occurred during the first millennium AD), mass migration of people – especially of Europe and Africa – for the Americas. During these periods can be discerned two categories of migration, i.e. spontaneous migration, at the initiative of individuals or smaller or larger groups (looking for resources fitting to their needs) and the second category, the forced migration of large masses of people (slave deportations) transformed into human capital. Since the nineteenth century the share of political and/or economic migration grows, decreasing almost all the second category.

These individuals movements are motivated by the economic attraction for new resources (land, hunting, natural wealth, easy access to capital etc.), to areas with high economic development, or with democratic opening, vis-à-vis the totalitarian doctrines, consistent with the aspirations of some degree of culture of the migrants. If for the Europeans in the XV-XVIII centuries, the migration was mainly spontaneous and had a clear motivation (rapid enrichment, release etc.), for the North African population is a completely different situation, in the sense that their position was not one of free individuals, but one of marketable labour resource (slavery). Forced migration resulting from marketing had as its main aim the best use of the recently discovered American lands and the maximizing of the profits of the new settlers.

Essentially being an economic type polarization, analytical approaches of the phenomenon by models acting as the attractive forces like gravity, electrostatic or, more generally, electromagnetic are understan-dable [18]. If for the gravity models, the attractors don't have a sign (positive, negative), in the Coulomb case the centres of influence have a sign and can explain the phenomenon of attraction, or rejection and from this perspective we consider it is a close description of actual economic situations.

As it is well known, the electric attractive forces are determined by the well known Coulomb's law [17, 18, 19]:

$$F = K \frac{q_1 q_2}{r_{12}^2} \qquad (4)$$

where $q_1$ and $q_2$ are the interacting electric charges of opposite sign (figure 1).

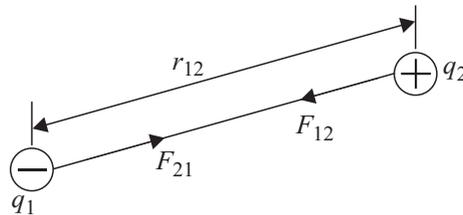

**Figure 1.** The electrical attraction between two electric charges of different signs.

where $r_{12}$ – is the distance between the two charges, and $K$ – is a proportionality constant, that depends on the properties of the environment between the two charges and of the chosen units system. The most common is the International System of Units (S.I.), where $K = \frac{1}{4\pi\varepsilon}$, for spherical symmetry and $K = \frac{1}{2\pi\varepsilon}$, for circular symmetry, $\varepsilon$ being the dielectric constant of the environment (electric permittivity).

For modelling the phenomenon of migration from neighbour regions (countries) *j*, with a powerful attraction characterized by a high socio-economic status (material resources, capital, developed logistics, information superstructures, banking, commercial or educational etc.) we will appeal to the notion of field (similar with the electric or electrostatic field) *E*, which is determined by the ratio of the power/force (electricity) *F* of the attractor and the charge $q_0$ on which acts the field [18]:

$$E = \frac{F}{q_0} \qquad (5)$$

In addition, the subject (individual or capital) can be subjected to the effects of a single centre or of multiple attractor centres, to a greater or lesser extent (attractive or repulsive).

Calculating the force of attraction on a charge $q_i$ symbolizing the number of emigrants from the region (country) $i$, we have:

$$F_a = q_i E \qquad (6)$$

or

$$F_a = \frac{q_i}{2\pi\varepsilon} \frac{Q_j}{R^2} \qquad (7)$$

where $Q_j = \sum_{1}^{n} {}_j q_j$ – is the charge distributed in the "rich region" $j$.

Knowing the intensity of the attraction forces (see eq. (7)), we can calculate the net migration flow given by $M_{ij}$ migrant's mass from the poor (negative) region (country) $i$, to the rich (positive) region (country) $j$:

$$M_{ij} = k \cdot F_a = k \frac{q_i \rho_j}{3\varepsilon} \frac{a^2}{R^2} \qquad (8)$$

where: $q_i$ – symbolizes the power of the region (country) of origin, $i$;

$\rho_j$ – symbolizes the power of the region (country) of destination, $j$;

$a$ – characterizes the sizes („radius") of region $j$;

$R$ – the distance between the two regions;

$k$ – the proportionality coefficient.

Using the relation (7) for the Coulomb type force $F_a$, the total mass of migrants $M_{ij}$ can also be written in the form:

$$M_{ij} = k \cdot F_a = k \frac{q_i}{2\pi\varepsilon} \frac{Q_j}{R^2} \qquad (9)$$

where $Q_j = \sum_{1}^{n} {}_j q_j$ symbolizes the economic power of the region (country) $j$ or all measurable in money resources (as PIB, e.g.).

In previous equations $\varepsilon$ represents the „environment's" permissiveness [1] between the two regions from our example $i$ and $j$.

---

[1] Term proposed by the authors, by similarity with the electromagnetic permittivity from physics.

As in the cases discussed in the previous subparagraph (v. §.2), can be written for the intensity of migration $M_{ij}$ equations like (1) ÷ (7) i.e.:

$$V_{ij} = R_{ij} - C_{ij} \tag{10}$$

or:

$$M_{ij} = f[P_i, P_j, D_{ij}, (U_j - U_i), (W_j - W_i)], \tag{11}$$

with the meanings of the terms expressed in these relations.

With the support of the new model, this kind of relations can even be explained for different concrete cases; e.g. the equalities between the relations (8) and (11) we can write:

$$k \frac{q_i}{2\pi\varepsilon} \frac{Q_j}{R^2} = f(P_i, P_j, D_{ij}, (U_j - U_i), (W_j - W_i)). \tag{12}$$

The physical quantities of the second member can be "correlated" to the physical quantities of the first member relations that include simultaneously electric variables $q_i$, $Q_j$ and economic variables: $(U_j - U_i)$, $(W_j - W_i)$, $P_i, P_j$ etc.

In the equations (8)-(11) from the model of electrostatic attraction forces, the „distance" costs $D_{ij}$ can be identified by the factor $k \frac{1}{2\pi\varepsilon} \frac{1}{R^2}$, where $R$ is the physical distance between the two regions.

### 4. Conclusions

Based on the analogy between the migration phenomena and electrostatic, electrical interactions has been proposed a new econophysics model for the migration phenomena. The new model has its limits in explaining the massive migration phenomena and the individual reasons of the immigrants, but at the same time it offers a summary of the phenomenon from a macroeconomic point of view and draws attention to the "desertification" of certain regions in qualified human resources, due to their attraction to more prosperous regions.

In addition we must take into account the following considerations:
• A possible change of the economic paradigm.
• The development of new communications and transactions via internet.
• The Knowledge Society could change the balance of economic power of the regions (countries).

# APPENDIX

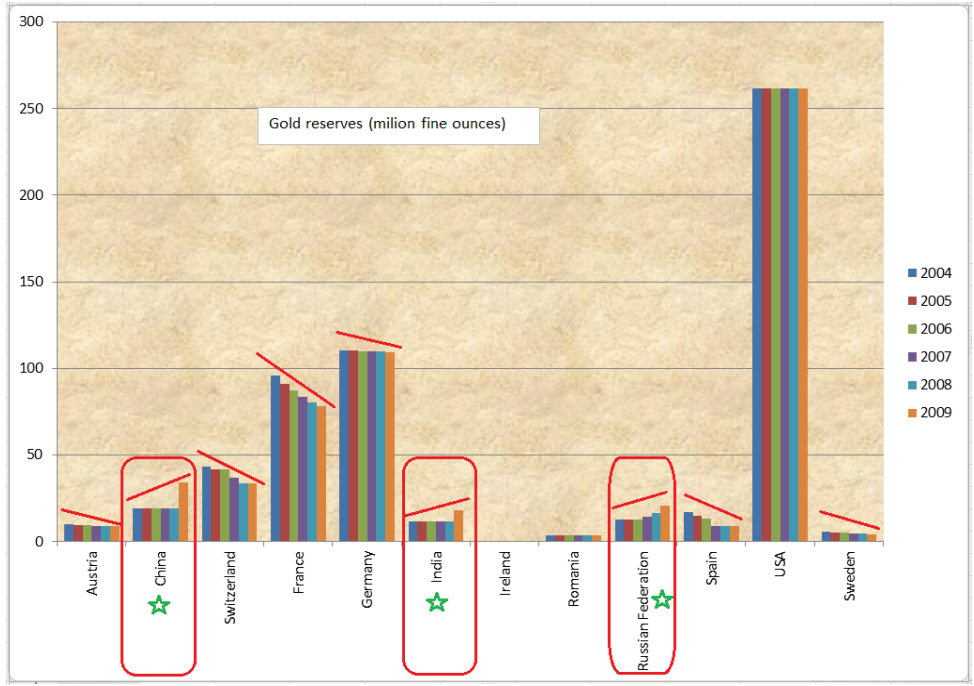

**Figure 2.** Gold Reserves (million fine ounces).

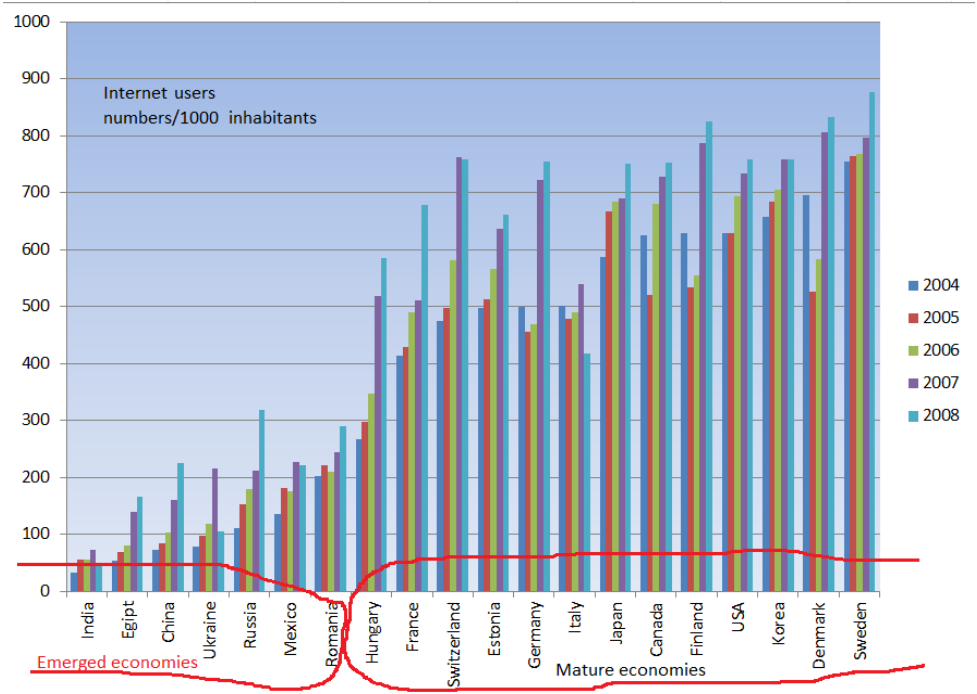

**Figure 3.** Internet users – numbers/1000 inhabitants.

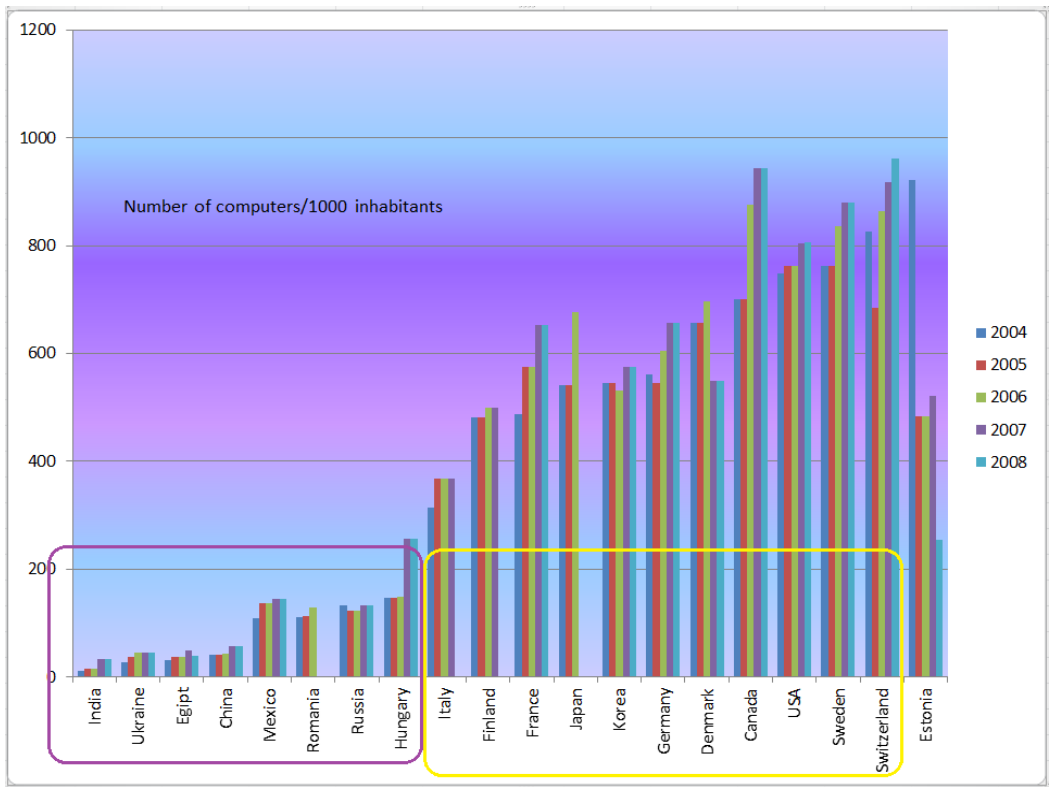

**Figure 4.** Number of computers/1000 inhabitants.

**Purchasing Power Standards (UE-27)**

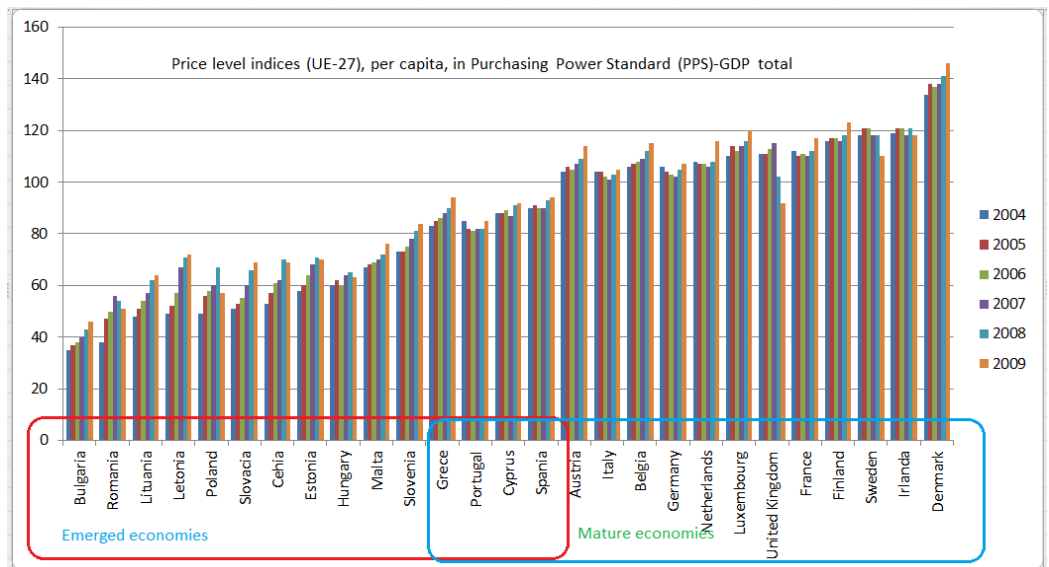

**Figure 5.** Price level indices (UE-27), per capita,
in Purchasing Power Standard (PPS)-GDP total.

283

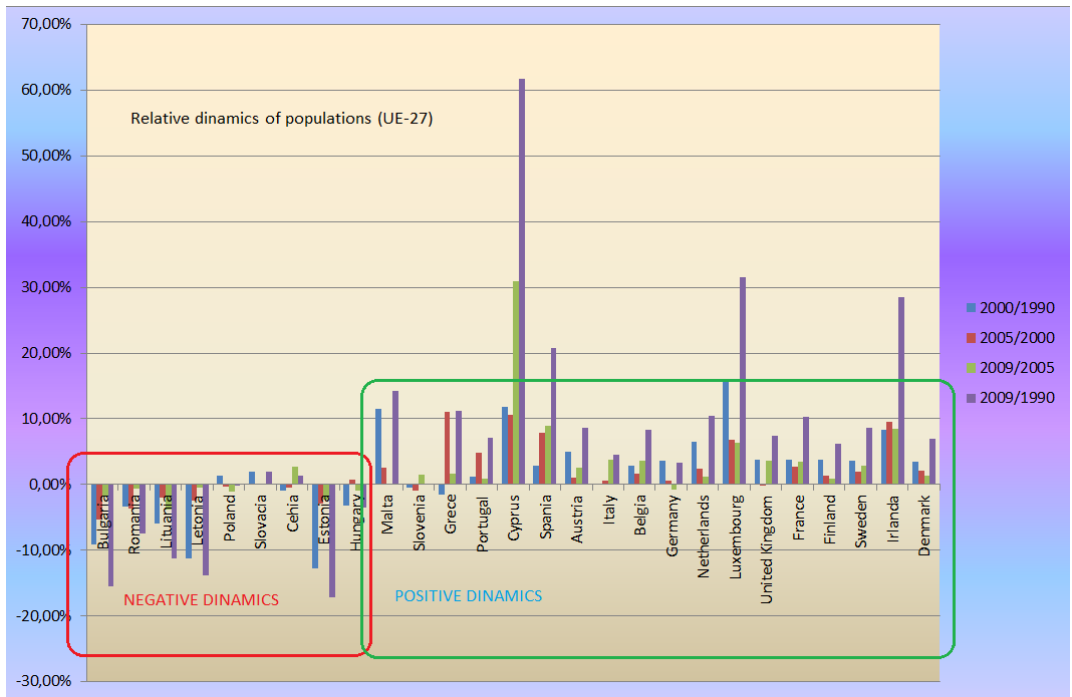

**Figure 6.** Relative dynamics of population (UE-27).